%% file: references.tex
\documentclass[review]{elsarticle}
\usepackage{hyperref}
\usepackage{graphicx}
\usepackage{tikz,pgfplots}
\usepackage{subfig}

\usepackage{pgf}
\usepackage[colorinlistoftodos]{todonotes}
\usepackage{caption}
\usepackage{array}

\definecolor{pinkk}{rgb}{1,0,1}
\definecolor{goldk}{rgb}{0.82,0.65,0.25}

\journal{Journal of \LaTeX\ Templates}

\begin{document} 

\begin{frontmatter}

\title{TF-IDFC-RF: A Novel Supervised Term Weighting Scheme for Sentiment Analysis}
\tnotetext[mytitlenote]{Fully documented templates are available in the elsarticle package on \href{http://www.ctan.org/tex-archive/macros/latex/contrib/elsarticle}{CTAN}.}

\author{Flavio Carvalho}
\author{Gustavo Paiva Guedes} 
\address{CEFET/RJ, 
Rio de Janeiro, Brazil}

\ead[url]{https://eic.cefet-rj.br/~gguedes/}

\author[mysecondaryaddress]{Computer Science Department\corref{mycorrespondingauthor}}
\cortext[mycorrespondingauthor]{Corresponding author}
\ead{gustavo.guedes@cefet-rj.br}



\begin{abstract}
 Sentiment Analysis is a branch of Affective Computing usually considered a binary classification task. In this line of reasoning, Sentiment Analysis can be applied in several contexts to classify the attitude expressed in text samples, for example, movie reviews, sarcasm, among others. A common approach to represent text samples is the use of the Vector Space Model to compute numerical feature vectors consisting of the weight of terms. The most popular term weighting scheme is TF-IDF (Term Frequency - Inverse Document Frequency). It is an Unsupervised Weighting Scheme (UWS) since it does not consider the class information in the weighting of terms.  Apart from that, there are Supervised Weighting Schemes (SWS), which consider the class information on term weighting calculation. Several SWS have been recently proposed, demonstrating better results than TF-IDF.
 In this scenario, this work presents a comparative study on different term weighting schemes and proposes a novel supervised term weighting scheme, named as TF-IDFC-RF. The effectiveness of TF-IDFC-RF is validated with SVM (Support Vector Machine) and NB (Naive Bayes) classifiers on four commonly used Sentiment Analysis datasets. TF-IDFC-RF shows promising results, outperforming all other weighting schemes on two datasets.

\end{abstract}


\end{frontmatter}


\section{Introduction}

Sentiment Analysis (SA) has attracted much attention in recent years \cite{Fang2015}.
It is a branch of Affective Computing usually considered a binary classification task \cite{PORIA201798}. 
The goal of SA is to classify the attitude expressed in text samples (e.g., positive or negative) rather than some facts (e.g., entertainment or sport) \cite{PORIA201798, cambria, Deng20143506}. It can be useful in several contexts, for example, to detect subjectivity \cite{pang-lee-2004-sentimental, Agrawal20181029}, irony/sarcasm  \cite{kumar2018sarcasm, filatova2012irony, Karoui2015644}, sentiment in movie reviews \cite{pang-lee-2004-sentimental,Kim2015775,Lee2018715}, among others.

A usual approach to represent text documents in the scope of SA is the use of the Vector Space Model (VSM), initially introduced in \cite{Salton1975613}. The main idea behind VSM is to represent each document as a numerical feature vector, consisting of the weight of terms extracted from the text corpus \cite{Chen20161339}. The weight of each term is considered the key component of document representation in VSM \cite{Deng20143506}. Thereby, the choice of the term weighting scheme to represent documents
directly affects the classification accuracy \cite{Domeniconi201639, Chen20161339, Dogan201945}.   

The weighting schemes can be divided into two main categories, based on the usage of class information in training documents \cite{Domeniconi201639}. The first one is the \textit{unsupervised term weighting} (UTW), which does not use class information to generate weights. The most popular unsupervised scheme is TF-IDF (Term Frequency - Inverse Document Frequency) \cite{Ren2013109, Gu2017436}. It has been used effectively in information retrieval studies; however, it is not very well suited for text classification tasks \cite{Dogan2019}. 
The second main category of weighting schemes is referred to as \textit{supervised term weighting} (STW), which was firstly proposed by Debole and Sebastiani \cite{debole2004supervised}.
STW schemes embrace class information from training dataset to compute term weighting \cite{Domeniconi201526}, which leads researchers to believe that these schemes have superior performance than UTW \cite{Lan2009721}.

Following this line of reasoning, several researches has focused on the development of new supervised weighting schemes (e.g., TF-RF \cite{Lan2009721}, TF-IDF-ICF \cite{Ren2013109}). Recently, Chen et al. \cite{Chen20161339} introduced a new term weighting based in inverse gravity moment, named as TF-IGM. The authors state that TF-IGM outperforms the state-of-the-art supervised term weighting schemes.
Dogan and Uysal \cite{Dogan201945} proposed TF-IGM$_{imp}$, which is an improved version of TF-IGM. Experiments indicated that TF-IGM$_{imp}$ outperforms TF-IGM. 

Although there are several supervised term weighting schemes, the experiments are usually conducted on multi-class datasets \cite{Nguyen201361, Chen20161339}. Based on this premise, Chen et al. \cite{Chen20161339} constructed ten two-class subsets from the Reuters-21578 corpus to conduct the experiments in their study. However,  Reuters-21578 is not an original two-class dataset, and the scope of this work is to study STW schemes in sentiment analysis, more specifically, in two-class datasets.

The main contributions of this work are: (i) the proposal of a novel STW scheme named as TF-IDFC-RF; (ii) the evaluation of ten weighting schemes (two UWS and eight SWS) on four two-class sentiment analysis datasets; we selected broadly available datasets to facilitate replication of the experiments.

According to the experimental results described in the next sections, TF-IDFC-RF outperforms all compared schemes in two datasets. These results are achieved in four two-class sentiment analysis datasets.

The remainder of this study is structured as follows: In Section \ref{sec:TWS} we present the concepts of nine STW schemes and the proposed scheme;  In Section \ref{sec:dataset}, we describe the datasets used in carrying out experiments; In Section \ref{sec:setup}, we discuss the experimental setup used to execute the experiments; In Section \ref{sec:experiments}, we present the experiments conducted with seven STW schemes, two UTW schemes and the proposed approach. Finally, in Section \ref{sec:conclusion}, we conclude and open discussion for further research.

\section{Term Weighting Schemes}
\label{sec:TWS}

As previously mentioned, VSM represents each document as a numerical feature vector (weights), where each dimension corresponds to a separate term (words, keywords, or longer phrases). The process of assigning a weight to each term is known as term weighting. There are several term weighting schemes in the literature, and the adoption of each of them leads to different results in text classification tasks \cite{Sabbah2017193, Tsai201114094}.

This section describes relevant concepts for text classification and discusses nine term weighting schemes in order to compare them with the proposed term weighting. Section \ref{subsec:utw}  reviews two of the most common unsupervised term weighting schemes, which are commonly considered as the baseline schemes. Section \ref{subsec:stw} presents seven supervised term weighting schemes used throughout this work.

\subsection{Unsupervised term weighting}
\label{subsec:utw}

Unsupervised term weighting schemes compute term weights considering information such as the frequency of terms in documents or the number of times that a term appears in a collection \cite{Sebastiani20021}.
In unsupervised term weighting approaches, the class information of the documents is not used to generate weights \cite{jin2005learn}. Section \ref{subsub:tf} and \ref{subsub:tfidf} present the UTW used in this work.

\subsubsection{Term Weighting Based on TF}
\label{subsub:tf}
Term frequency (TF) is the number of times a particular term $t_i$ occurs in a document $d_j$, as indicated in Eq.~\ref{tf}. It is one of the most important term weighting schemes in document analysis \cite{Li201366}. However, it is widely recognized that TF puts too much weight on repeated occurrences of a term \cite{Lv20117}.

\begin{equation}
W_{TF}(t_i)=TF(t_i,d_j) 
\label{tf}
\end{equation}

\subsubsection{Term Weighting Based on TF-IDF}
\label{subsub:tfidf}

TF-IDF \cite{Salton1975613} is one of the earliest and common unsupervised weighting methods \cite{Dogan2019}. 
The intuition behind TF-IDF is that, for some context, some terms are more important than others to describe documents. For example, a term that appears in all documents does not have substantial relevance to help identifying documents. Eq.~\ref{tfidf} describes TF-IDF, where  $N$ is the number of documents in the corpus and $DF(t_i)$ corresponds to the frequency of documents that term $t_i$ appears in the collection. TF-IDF and TF are considered unsupervised term weighting schemes as they do not take into account class information.

\begin{equation}
W_{TF.IDF}(t_i)=TF(t_i,d_j) \times \log\left( \frac{N}{DF(t_i)} \right)
\label{tfidf}
\end{equation}

\subsection{Supervised term weighting}
\label{subsec:stw}

This section describes seven supervised term weights schemes and the proposed weighting scheme.
As already stated, STW schemes weight terms by exploiting the known class information in training corpus. The fundamental elements of supervised term weighting are depicted in Table \ref{tab:fundamentals}.

\begin{table}[htb!]
\centering
\setlength{\tabcolsep}{2em}
\begin{tabular}{ccc}
\hline
& $c_k$ & $\overline{c_k}$ \\
\hline
$t_i$& A & C \\
$\overline{t_i}$& B & D \\
\hline
\end{tabular}
\caption{Notation for supervised term weighting schemes.}
\label{tab:fundamentals}
\end{table}

 In this representation, the importance of a term $t_i$ for a class $c_k$ is represented as follows:
 
\begin{itemize}
    \item A represents the number of documents in class $c_k$ where the term $t_i$ occurs at least once;
    \item C represents the number of documents not belonging to class $c_k$ where the term $t_i$ occurs at least once;
    
    \item B represents the number of documents belonging to class $c_k$ where the term $t_i$ does not occur;
    
    \item D represents the number of documents not belonging to class $c_k$ where the term $t_i$ does not occur;
    
    \item N is the total number of documents in the corpus; N = A + B + C + D;
\item N$_p$ is the number of documents in the positive class;
N$_p$ = A + B;
\item N$_n$ is the number of documents in the negative class; N$_n$ = C + D.
\end{itemize}

\subsubsection{Term Weighting Based on Delta TF-IDF}

Delta TF-IDF was proposed by Martineau and Finin \cite{MartineauF09}. It computes the difference of TF-IDF
scores in the positive and negative classes to improve accuracy \cite{MartineauF09}.
As an STW, it considers the distribution of features between the two classes before classification,  recognizing and heightening the effect of distinguishing terms. Delta TF-IDF boosts the importance of words that are unevenly distributed between the positive and the negative class. 

 In this work, we use the smoothed version, as indicated in Eq.~\ref{deltaidf}, since it achieved higher accuracy in Paltoglou and Thelwall \cite{Paltoglou20101386}. N$_{p}$ and N$_{n}$ are, respectively, the number of documents in positive and negative classes. A and C represent the document frequency of term $t_i$ in positive and negative classes, respectively. 

\begin{equation}
W_{\delta.TF.IDF}(t_i)=TF(t_i,d_j) \times log_2\left( \frac{N_{p}  \times C + 0.5}{A \times N_{n} + 0.5} \right)
\label{deltaidf}
\end{equation}

\subsubsection{Term Weighting Based on TF-IDF-ICF}

TF-IDF-ICF is a supervised weighting scheme based on traditional TF-IDF. However, it adds \textit{Inverse Class Frequency (ICF)} factor \cite{Ren2013109} to give higher weighting values to rare terms that occur in fewer documents (IDF) and classes (ICF). In Eq.~\ref{icf}, $M$ is the number of classes in the collection and $CF(t_i)$ corresponds to the frequency of classes that term $t_i$ appears in the collection.

\begin{equation}
W_{TF.IDF.ICF}(t_i)=TF(t_i,d_j) \times IDF(t_i) \times \left(1 + log\left( \frac{M}{CF(t_i)} \right)\right)
\label{icf}
\end{equation}

\subsubsection{Term Weighting Based on TF-RF}

TF-RF (\textit{Term Frequency - Relevance Frequency}) was proposed in \cite{Lan2009721}. Similar to Delta TF-IDF, TF-RF takes into account terms distribution in positive and negative classes. However, only the documents containing the term are considered, that is, the Relevance Frequency (RF) of the terms. TF-RF is indicated in Eq. \ref{tfrf}, where the minimal denominator is $1$ to avoid division by zero.

\begin{equation}
W_{TF.RF}(t_i) = TF(t_i,d_j) \times log_2 \left(2+\frac{A}{max(1,C)} \right)
\label{tfrf}
\end{equation}

\subsubsection{Term Weighting Based on TF-IGM}

Term Frequency - Inverse Gravity Moment (TF-IGM) \cite{Chen20161339}
is proposed to measure the non-uniformity or concentration of terms inter-class distribution, which reflects the terms class distinguishing power. 
The standard IGM equation assign ranks ($r$) based on the inter-class distribution concentration of a term, which is analogous to the concept of ``gravity moment'' (GM) from the physics. IGM is indicated in Eq. \ref{igm}, where ${f_{ir}}$ $(r = 1, 2, ..., M)$ indicates the number of documents containing the term ${t_i}$ in the $r$-th class, which are sorted in descending order. Thus, $f_{i1}$  represents the frequency of $t_i$ in the class which it appears most often. 

\begin{equation}
{IGM}(t_i)= \left ( \frac{f_{i1}}{\sum_{r=1}^{M}{f_{ir}}\times r} \right )
\label{igm}
\end{equation}

{TF-IGM} term weighting is then defined based on IGM($t_i$), as shown in Eq. \ref{tfigm}. $\lambda$ is an adjustable coefficient used to maintain the relative balance between
the global and local factors in the weight of a term.
The $\lambda$ coefficient has a default value of $7.0$ and can be set as a value between $5.0$ and $9.0$ \cite{Chen20161339}. Eq. \ref{stfigm} presents SQRT\_TF-IGM, which calculates the square root of TF, as a technique to obtaining a more reasonable term weighting by reducing the effect of high TF \cite{Chen20161339}.

\begin{equation}
W_{TF.IGM}(t_i)=  TF(t_i,d_j)  \times \left ( 1 + \lambda \times  {IGM}(t_i) \right )
\label{tfigm}
\end{equation}

\begin{equation}
W_{SQRT\_TF.IGM}(t_i)=  SQRT\_TF(t_i,d_j)  \times \left ( 1 + \lambda \times  {IGM}(t_i) \right )
\label{stfigm}
\end{equation}

To enhance the weighting process of TF-IGM for extreme scenarios, Dogan and Uysal \cite{Dogan201945} proposed IGM$_{imp}$, an improvement of ${IGM}$. {IGM$_{imp}$} is used in two new term weighting schemes, {TF-IGM$_{imp}$} and {SQRT\_TF-IGM$_{imp}$}, which were also proposed in \cite{Dogan201945}. IGM$_{imp}$ is described in Eq. \ref{igmimp}, where $D_{total}  \left (t_{i\_max}\right )$
indicates the total number of documents in the class that the term $t_i$ occurs most. TF-IGM$_{imp}$ and SQRT\_TF-IGM$_{imp}$ are defined in Eq.~\ref{tfigmimp} and Eq.~\ref{stfigmimp}, respectively. Dogan and Uysal \cite{Dogan201945} report IGM$_{imp}$ produces better results than IGM.

\begin{equation}
{IGM.{imp}}(t_i) =  \left [ \frac{f_{i1}}{\sum_{r=1}^{M}{f_{ir}}\times r + log_{10}\left ( \frac{D_{total}  \left (t_{i\_max}\right )}{f_{i1}} \right )} \right ]
\label{igmimp}
\end{equation}

\begin{equation}
W_{TF.IGM_{imp}}(t_i) =  TF(t_i,d_j)  \times \left ( 1 + \lambda \times {IGM_{imp}}(t_i) \right )
\label{tfigmimp}
\end{equation}

\begin{equation}
W_{SQRT.TF.IGM_{imp}}(t_i) =  SQRT\_TF(t_i,d_j)  \times \left ( 1 + \lambda \times {IGM_{imp}}(t_i) \right )
\label{stfigmimp}
\end{equation}

\subsection{Novel Term Weighting Scheme}

The proposed term weighting scheme is based on the IDF concept. However, it calculates the inverse document frequency of terms in classes (IDFC). It is also inspired in TF-RF since it calculates the Relevance Factor of a term. 

Eq.~\ref{tfeu} describes the proposed supervised term weighting scheme, named as TF-IDFC-RF. To avoid division by zero, we adjust the denominators with (A + 1) for IDFC and (C + 1) for RF as in \cite{Lan2009721}. In the RF part, we also adjust the numerator with (A + 1) to  avoid $\log (0)$.

\begin{equation}
W_{TF.IDFC.RF}(t_i) = SQRT\_TF(t_i,d_j) \times
log_2\Big(\frac{2+max(A,C)}{max(2, min(A,C))}\Big) \times \sqrt{B+D}
\label{tfeu}
\end{equation}


To illustrate the properties of different term weighting measures and to obtain a more solid understanding of TF-IDFC-RF, consider the fundamental elements presented in Table \ref{tab:fundamentals}.
Suppose a training dataset containing $100$ documents. Consider now the distribution of terms $t_1$ and $t_2$ for two classes $c_p$ and $c_n$, as defined in Table \ref{tab:fundamental2}\footnote{In our case, $c_n$ corresponds to $\overline{c_k}$, since we are focused in the two-class problem}.

\begin{table}[htb!]
\centering
\begin{tabular}{ccc|c|ccc}
\hline
& $c_p$ & $c_n$ & & & $c_p$ & $c_n$  \\
\hline
$t_1$& 27 & 5 & &$t_2$& 10 & 25\\
$\overline{t_1}$& 3 & 65 & & $\overline{t_2}$& 20 & 45\\
\hline
\end{tabular}
\caption{Example of document distribution for two terms.}
\label{tab:fundamental2}
\end{table}

Taking into account the $t_1$ distribution in Table \ref{tab:fundamental2}, the weighting calculation for IDF, Delta IDF, IDF-ICF, RF, TF-IDF-RF, IGM and IGM$_{imp}$ is as follows:

\begin{table}[htb!]
\small
\centering
\begin{tabular}{ll}
\hline
Weighting Scheme & Calculation\\
\hline
IDF($t_1$)&log(100/(27+5))=log(3.125) = 0.4949\\

Delta IDF($t_1$) &log$_2$((30*5+0.5)/(27*70+0.5))=-3.6510\\
IDF-ICF($t_1$)&(1+0.4949)*(1+log(2/2))=1.4949\\
RF($t_1$) & log$_2$(2+(27/5))=2.8875\\
IGM($t_1$)&27/((27*1)+(5*2))=0.7297\\
IGM$_{imp}$($t_1$)&27/((27*1)+(5*2)+0.0458)=0.7288\\
IDFC-RF($t_1$)&$log_2$((2+27)/5)*8.2462)=20.9128\\
  \hline
\end{tabular}
\caption{Scores of term weighting schemes considering distribution in Table \ref{tab:fundamental2}.}
\label{tab:scores}
\end{table}

In order to investigate the effect produced by these schemes, Table \ref{tab:scores2} summarizes the scores for both terms $t_1$ and $t_2$. When comparing IDFC-RF with RF, it is possible to note that IDFC-RF seems to be less discriminative between the terms. For example, the ratio between the terms (i.e., $t_1$ and $t_2$) is less prominent in IDFC-RF. Therefore, IDFC-RF considers intra-class and inter-class distribution since both are even taken as equally important in STW \cite{Chen20161339}. 


\begin{table}[htb!]
\centering
\begin{tabular}{lll}
\hline
Weighting Scheme & $t_1$ &  $t_2$ \\
\hline
  IDF&0.4949&1.0498\\
  Delta IDF&-3.6510&0.0995\\
  IDF-ICF&1.4949&1.0498\\
  RF&2.8875&1.2630\\
  IGM&0.7297&0.5556\\
  IGMimp&0.7288&0.5501\\
  IDFC-RF&20.9128&21.7681\\
  \hline
\end{tabular}
\caption{Scores of term weighting schemes considering distribution in Table \ref{tab:fundamental2}.}
\label{tab:scores2}
\end{table}

\section{Datasets} 
\label{sec:dataset}
This section describes the datasets used to produce the experiments. All datasets are commonly used in
Sentiment Analysis studies.

\subsection{Polarity}
\label{subsec:pol}
The Polarity dataset consists of $1,000$ positive and $1,000$ negative movie reviews. It was first introduced by Pang and Lee \cite{pang-lee-2004-sentimental}. It is used as a baseline dataset in several sentiment analysis.

\subsection{Amazon Sarcasm}
\label{subsec:sarcasm}
The Amazon Sarcasm dataset was introduced by Filatova \cite{filatova2012irony}. It consists of an unbalanced dataset with $437$ sarcastic reviews and $817$ regular reviews from Amazon (\url{http://www.amazon.com}). The reviews were labeled using crowdsourcing. 

\subsection{Subjectivity}
\label{subsec:sub}
The Subjectivity dataset was introduced by
Pang and Lee \cite{pang-lee-2004-sentimental}. It consists of $5,000$ subjective sentences and $5,000$ objective sentences. The subjective sentences were collected from \url{www.rottentomatoes.com}. The objective sentences were extracted from  summaries available from the Internet Movie Database (\url{www.imdb.com}).

\subsection{Movie Review}
\label{subsec:mr}
Movie Review dataset contains $10,662$ movie-reviews ``snippets'' (a striking extract usually one sentence long) with positive and negative labels \cite{pang2005seeing}. The movie-reviews were collected from \url{www.rottentomatoes.com}. It consists of $5,331$ negative snippets and $5,331$ positive snippets. 
\section{Experimental setup}
\label{sec:setup}
This section describes the experimental setup used to present the experimental results. In Section \ref{subsec:classific} we discuss the classification process adopted in this work. In Section \ref{subsec:learning} we review concepts from the learning algorithms considered to produce the experiments. Finally, in Section  \ref{subsec:performanceM} we describe the evaluation measures used in the experimental study.

\subsection{Classification Process}
\label{subsec:classific}

\begin{figure}[hbt!]
 \centering
\includegraphics[width=12.0cm]{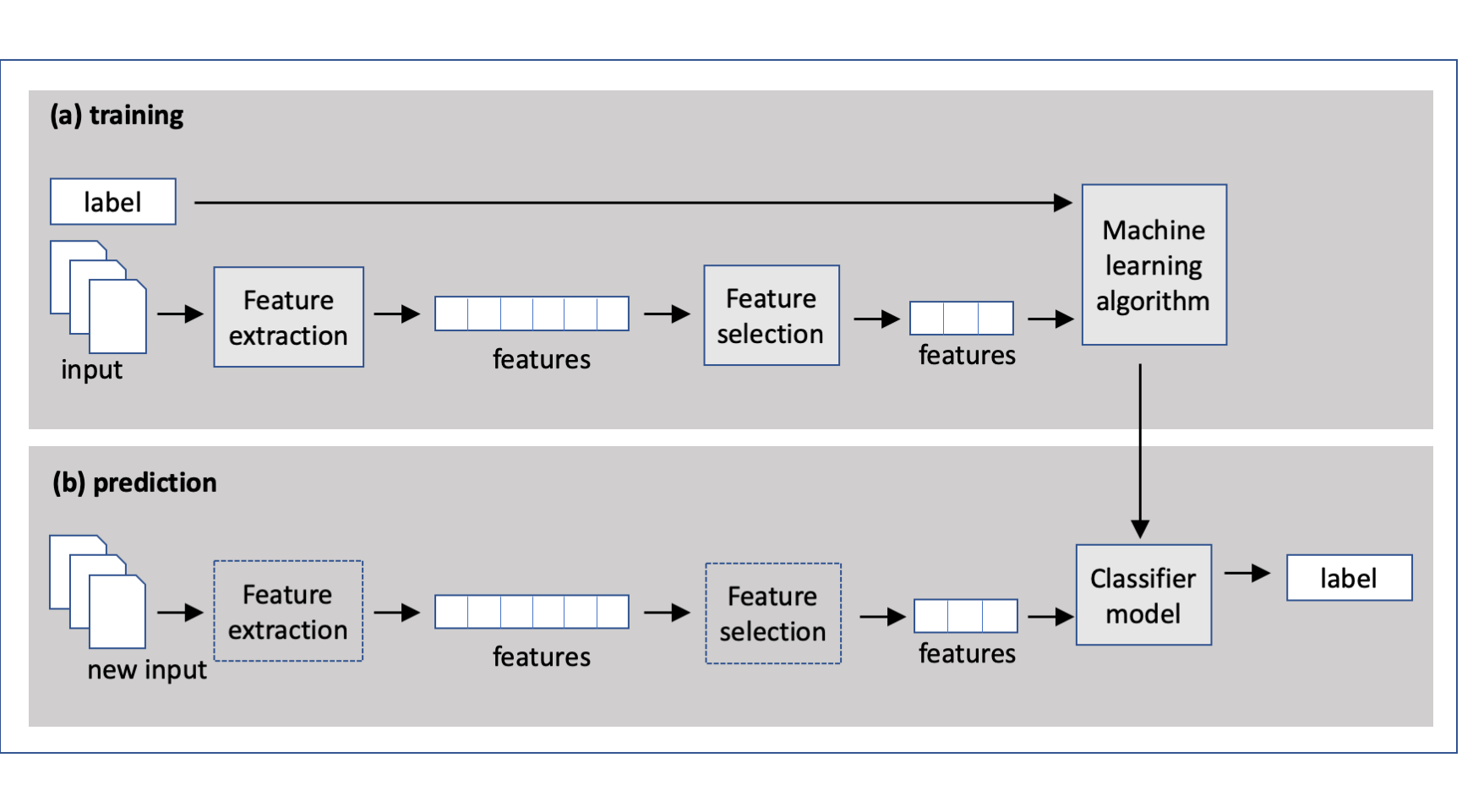}
\caption{Supervised classification process. Adapted from He \cite{he2013text}. }
\label{fig-classification}
\end{figure}

All documents in each dataset were preprocessed with lowercase conversion,  punctuation removal, and tokenization. 
In the classification process, we applied the stratified 5-fold cross-validation technique to present the classification performance.
The process adopted to execute the experiments is based on the training-and-testing paradigm described in \cite{jurafsky2008speech}.
The procedure followed for each fold is illustrated in Fig. \ref{fig-classification}. As depicted in Fig.~\ref{fig-classification}(a), during the training phase, a feature extraction step (i.e., a term weighting scheme) helps to convert each text into a feature vector. This step can include a feature selection method to reduce the feature set size. Finally, the feature set is fed into a machine-learning algorithm to generate a model. As depicted in Fig.~\ref{fig-classification}(b), during prediction, the statistical parameters of the training set  (i.e., the classifier model generated in training phase) are used to compute the features of unseen inputs (Feature identification) \cite{jurafsky2008speech}. These feature sets are then fed into the model to generate the output labels.

\textbf{Feature extraction: } In the scope of this study, in the ``feature extraction'' step, we extracted the features with the following schemes: TF-IDFC-RF with TF, TF-IDF, Delta TD-IDF (DTF-IDF), TF-IDF-ICF, TF-RF, TF-IGM, SQRT\_TF-IGM (STF-IGM), TF-IGM$_{imp}$, and SQRT\_ST-IGM$_{imp}$ (STF-IGM$_{imp}$). Next, as in \cite{Chen20161339, Dogan201945}, we adopted  $\chi^2$ or chi-square statistics (CHI2)  for feature selection\footnote{For more information about feature selection methods, please refer to \cite{yang1997comparative}.}. 
The weighting schemes are tested with top {500, 1,000, 2,000, 4,000, 6,000, 8,000, 10,000, 12,000, and 14,000} terms scored and sorted descending order by $CHI2_{max}$ for all datasets.
\textbf{Parameters setting: } Lambda parameters were configured with $\lambda=7$ for TF-IGM, STF-IGM, TF-IGM$_{imp}$ and STF-IGM$_{imp}$, as it is considered the default value \cite{Chen20161339}. 


\subsection{Learning Algorithms}
\label{subsec:learning}
To evaluate the effectiveness of weighting schemes, we conducted the experiments with Support Vector Machines (SVM), since it is the best learning approach in text categorization \cite{Sebastiani20021, Lan2009721, Domeniconi201526}. We used the scikit-learn implementation of SVC \cite{sklearn_api} trained with default values.

We also executed the experiments with the Naive Bayes algorithm (NB), since it is also often used as a baseline for text categorization and sentiment analysis \cite{Wang201290}. We also used a Sklearn implementation of NB, named as MultinomialNB.

\subsection{Performance Measures}
\label{subsec:performanceM}
We calculated the effectiveness of the weighting schemes using weighted F$_1$ measure, as described in Chavarriaga et al. \cite{Chavarriaga20132033}. 
 The weighted F$_1$ measure
is calculated considering the class size and the \textit{precision} and \textit{recall} for each class.
\textit{Precision} is defined as the fraction of all positive predictions that are actual positives, as defined in Eq. \ref{precision}. \textit{Recall} is the fraction of all actual positives that are predicted to be positive, as indicated in Eq. \ref{recall}.

\begin{equation}
    precision = \frac{TP}{TP+FP}
    \label{precision}
\end{equation}

\begin{equation}
    recall = \frac{TP}{TP+FN}
    \label{recall}
\end{equation}

Considering the two equations above, Weighted F$_{1}$ measure is defined as the Eq. \ref{F$_1$}.

\begin{equation}
F_{1}=\sum_{i}2*w_{i}*\frac{precision_{i}\cdot recall_{i}}{precision_{i} + recall_{i}}
\label{F$_1$}
\end{equation}

 In Eq. \ref{F$_1$}, $i$ is the class index and $w_i = n_i/N$ is the proportion of samples of class $i$. $N$ indicates the total number of samples and $n_i$ denotes the number of samples of the $i^{th}$ class.


\section{Results}
\label{sec:experiments}

In this section, we evaluate the performance of the unsupervised term weighting scheme proposed in this work, named as TF-IDFC-RF. To accomplish this goal, we compared TF-IDFC-RF with $9$ other weighting schemes on \textit{Polarity}, \textit{Amazon Sarcasm}, \textit{Subjectivity} and\textit{ Movie Review} datasets.









\subsection{Performance comparisons on the Polarity dataset}

\begin{figure}[htb!]
\centering
    \centering
    \subfloat[NB]{\resizebox{0.58\textwidth}{!}{\input{fig2a.tex}}}%
    \subfloat[SVM]{\resizebox{0.42\textwidth}{!}{\input{fig2b.tex}}}%
\caption{Weighted-F$_1$ scores obtained using NB and SVM classifiers with $10$ term weighting schemes on \textit{Polarity} dataset.}
\label{pol2}
\end{figure}
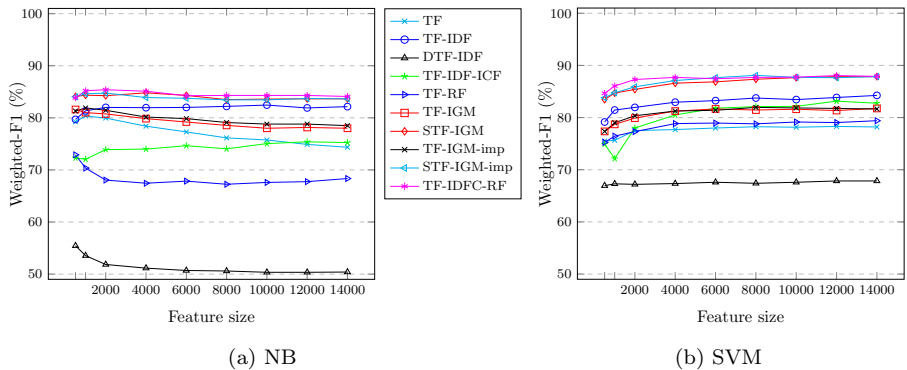

Figures~\ref{pol2}(a) and \ref{pol2} (b) report the Weighted F$_1$ score obtained with NB and SVM on \textit{Polarity dataset} considering $10$ different term weighting schemes. It is important to note that this dataset is balanced, as described in Section \ref{subsec:pol}.
TF-IDFC-RF shows the best performance concerning both classifiers with most variations of feature size.
STF-IGM presents the second-best performance with NB, especially when the feature sizes are $4,000$ and $6,000$. STF-IGM$_{imp}$ has the second-best results with SVM classifier, specifically when the feature size is $6,000$ and $8,000$. Generally, the squared versions of IGM (STF-IGM and STF-IGM$_{imp}$) show better results than the non-squared versions.
DTF-IDF presents the worst result with both classifiers. TF-RF and TF-IDF-ICF also produced poor results. The UTW schemes (TF and TF-IDF) results are lower than TF-IDFC-RF concerning all feature sizes.


Tables \ref{tab:polnb} and \ref{tab:polsvm} describe in detail the F$_1$ score achieved with NB and SVM classifiers with Polarity dataset. It is possible to note an increase in the performance of TF-IDFC-RF when the feature size also increases.  
The superiority of TF-IDFC-RF over IGM based schemes is more evident with SVM classifier. For example, when feature sizes are between $500$ and $2000$, the difference between both algorithms achieves more than one percentage point.

\begin{table}[htb!]
\footnotesize
\centering
\setlength{\tabcolsep}{0.58em}
\begin{tabular}{lllllllllll}
\hline
Feat.  &  \multicolumn{3}{l}{\textit{Weighted-F$_1$} (\%)}   &   &  & \\
\cline{2-11}
Size &  TF& TF-&  DTF- &  TF- & TF-&  TF-& STF-& TF-&  STF- &TF-\\
 &  & IDF&  IDF &  ICF & RF&  IGM& IGM& IGMi& IGMi &IDFC-RF\\
\hline

500 & 79.25 & 79.70 & 55.45 & 72.30 & 72.90 & 81.60 & 84.10 & 81.30 & 84.05 & \textbf{84.25} \\
1000 & 80.35 & 81.35 & 53.55 & 72.05 & 70.35 & 81.00 & 84.35 & 81.90 & 84.60 & \textbf{84.95} \\
2000 & 79.95 & 82.00 & 51.85 & 73.90 & 68.05 & 80.75 & 84.25 & 81.45 & \textbf{84.75} & \textbf{84.75} \\
4000 & 78.40 & 81.95 & 51.15 & 74.00 & 67.45 & 79.85 & \textbf{84.80} & 80.15 & 83.90 & 84.50 \\
6000 & 77.30 & 82.00 & 50.70 & 74.65 & 67.85 & 79.20 & \textbf{84.30} & 79.85 & 83.75 & 83.65 \\
8000 & 76.15 & 82.20 & 50.60 & 74.05 & 67.25 & 78.55 & 83.50 & 79.10 & 83.45 & \textbf{83.80} \\
10000 & 75.75 & 82.45 & 50.35 & 75.05 & 67.60 & 78.00 & 83.60 & 78.80 & 83.45 & \textbf{83.95} \\
12000 & 74.90 & 81.90 & 50.35 & 75.40 & 67.75 & 78.15 & 83.65 & 78.80 & 83.70 & \textbf{84.00} \\
14000 & 74.35 & 82.15 & 50.40 & 75.25 & 68.35 & 78.00 & 83.65 & 78.50 & 83.65 & \textbf{84.05} \\

\hline
\end{tabular}
\caption{Performances of TF,  TF-IDF,  DTF-IDF, TF-IDF-ICF,  TF-RF,  TF-IGM,  STF-IGM,  TF-IGM$_{imp}$,  STF-IGM$_{imp}$, and  TF-IDFC-RF using NB classifier on \textit{Polarity} dataset.}
\label{tab:polnb}
\end{table}

\begin{table}[htb!]
\footnotesize
\centering
\setlength{\tabcolsep}{0.58em}
\begin{tabular}{lllllllllll}
\hline
Feat.  &  \multicolumn{3}{l}{\textit{Weighted-F$_1$} (\%)}   &   &  & \\
\cline{2-11}
Size &  TF& TF-&  DTF- &  TF- & TF-&  TF-& STF-& TF-&  STF- & TF-\\
 &  & IDF&  IDF & ICF & RF&  IGM& IGM& IGMi& IGMi & IDFC-RF \\
\hline

500 & 75.50 & 79.15 & 66.95 & 74.90 & 75.25 & 77.35 & 83.45 & 77.35 & 83.80 & \textbf{85.05} \\
1000 & 75.65 & 81.45 & 67.30 & 72.15 & 76.45 & 78.70 & 84.70 & 79.00 & 84.75 & \textbf{85.95} \\
2000 & 77.50 & 81.95 & 67.20 & 78.00 & 77.30 & 79.90 & 85.40 & 80.35 & 85.90 & \textbf{87.05} \\
4000 & 77.70 & 82.95 & 67.35 & 80.50 & 78.85 & 81.25 & 86.60 & 81.25 & 87.10 & \textbf{87.90} \\
6000 & 78.00 & 83.25 & 67.60 & 81.80 & 78.95 & 81.65 & 86.85 & 81.40 & \textbf{87.65} & 87.30 \\
8000 & 78.25 & 83.75 & 67.40 & 82.10 & 78.80 & 81.45 & 87.35 & 81.95 & \textbf{88.10} & 87.55 \\
10000 & 78.15 & 83.45 & 67.60 & 82.15 & 79.15 & 81.60 & 87.65 & 81.85 & 87.70 & \textbf{87.85} \\
12000 & 78.30 & 83.85 & 67.85 & 83.15 & 79.05 & 81.35 & 87.85 & 81.75 & 87.65 & \textbf{87.85} \\
14000 & 78.20 & 84.25 & 67.85 & 82.75 & 79.40 & 81.75 & 87.85 & 81.65 & 87.85 & \textbf{88.30} \\

\hline
\end{tabular}
\caption{Performances of TF,  TF-IDF,  DTF-IDF, TF-IDF-ICF,  TF-RF,  TF-IGM,  STF-IGM,  TF-IGM$_{imp}$,  STF-IGM$_{imp}$, and  TF-IDFC-RF using SVM classifier on \textit{Polarity} dataset.}
\label{tab:polsvm}
\end{table}

\subsection{Performance comparisons on the Sarcasm dataset}

\begin{figure}[htb!]
\centering
    \centering
    \subfloat[NB]{\resizebox{0.58\textwidth}{!}{\input{fig3a.tex} }}%
    \subfloat[SVM]{\resizebox{0.42\textwidth}{!}{\input{fig3b.tex}}}%
\caption{Weighted-F$_1$ scores obtained using NB and SVM classifiers with $10$ term weighting schemes on \textit{Sarcasm} dataset.}
\label{fig:irony}
\end{figure}
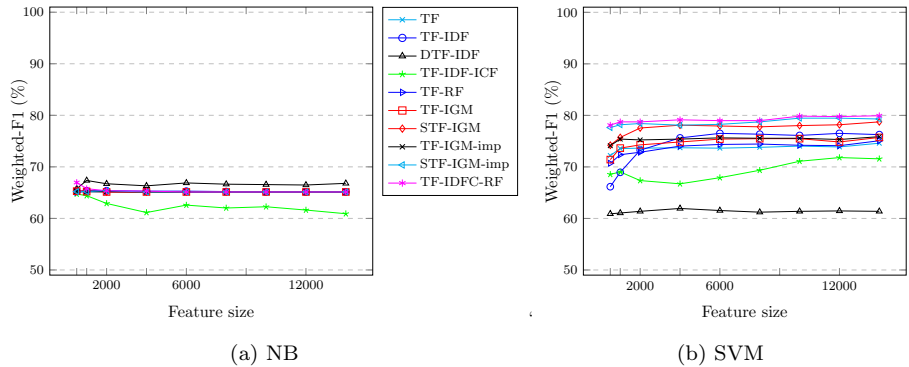

Figures \ref{fig:irony}(a) and \ref{fig:irony}(b) present Weighted F$_1$ scores achieved with NB and SVM classifiers on Sarcasm dataset, which is an unbalanced small dataset, as described in Section \ref{subsec:sarcasm}. TF-IDFC-RF produced the best performance with SVM classifier. DTF-IDF presented the best results with NB classifier and the worst results with SVM classifier.  STF-IGM$_{imp}$ is the second-best weighting scheme considering the SVM classifier. TF-RF also showed poor results with SVM. 

Detailed results on Sarcasm dataset are reported in Tables \ref{tab:sarcNB} and \ref{tab:sarcSMO}. It is evident the superiority of TF-IDFC-RF over all other weighting schemes with SVM. DTF-IDF has the best overall performance with NB, reaching an F$_1$ score of $66.80\%$  with $14,000$ features, as indicated in Table \ref{tab:sarcNB}. With the same feature size, TF-IDFC-RF achieves an F$_1$ score of $65.13\%$. It is clear that when discussing the performance of TF-IDFC-RF and NB, it is not so meaningful as the result reported by the TF-IDFC-RF and SVM, which achieved $79.93\%$ of $F_1$ with $14,000$ features. However, TF-IDFC-RF presents the second-best result with NB classifier.

Although TF-IDFC-RF was inferior to DTF-IDF considering the NB classifier, it is possible to notice that all F$_1$ values achieved by TF-IDFC-RF and SVM are higher than the values reported by DTF-IDF and NB. This evidence points out that, when considering the Sarcasm dataset,  TF-IDFC-RF obtains better results than DTF-IDF.


\begin{table}[htb!]
\footnotesize
\centering
\setlength{\tabcolsep}{0.58em}
\begin{tabular}{lllllllllll}
\hline
Feat.  &  \multicolumn{3}{l}{\textit{Weighted-F$_1$} (\%)}   &   &  & \\
\cline{2-11}
Size &  TF& TF-&  DTF- &  TF- & TF-&  TF-& STF-& TF-&  STF- & TF-\\
 &  & IDF&  IDF &  ICF & RF&  IGM& IGM& IGMi& IGMi & IDFC-RF\\
\hline
500 & 65.21 & 65.29 & 65.76 & 64.73 & 65.05 & 65.29 & 65.21 & 65.21 & 65.21 & \textbf{66.96} \\
1000 & 65.13 & 65.37 & \textbf{67.36} & 64.41 & 65.29 & 65.21 & 65.37 & 65.21 & 65.21 & 65.76 \\
2000 & 65.13 & 65.37 & \textbf{66.72} & 62.90 & 65.21 & 65.13 & 65.13 & 65.21 & 65.37 & 65.37 \\
4000 & 65.13 & 65.29 & \textbf{66.32} & 61.14 & 65.21 & 65.13 & 65.13 & 65.13 & 65.13 & 65.29 \\
6000 & 65.13 & 65.29 & \textbf{66.88} & 62.58 & 65.13 & 65.13 & 65.13 & 65.13 & 65.13 & 65.13 \\
8000 & 65.13 & 65.21 & \textbf{66.64} & 62.02 & 65.13 & 65.13 & 65.13 & 65.13 & 65.13 & 65.13 \\
10000 & 65.13 & 65.13 & \textbf{66.56} & 62.26 & 65.13 & 65.13 & 65.13 & 65.13 & 65.13 & 65.13 \\
12000 & 65.13 & 65.13 & \textbf{66.48} & 61.62 & 65.13 & 65.13 & 65.13 & 65.13 & 65.13 & 65.13 \\
14000 & 65.13 & 65.13 & \textbf{66.80} & 60.91 & 65.13 & 65.13 & 65.13 & 65.13 & 65.13 & 65.13 \\

\hline
\end{tabular}
\caption{Performances of TF, TF-IDF, DTF-IDF, TF-IDF-ICF, TF-RF, TF-IGM, STF-IGM, TF-IGM$_{imp}$, STF-IGM$_{imp}$ and TF-IDFC-RF using NB classifier on \textit{Sarcasm} dataset.}
\label{tab:sarcNB}
\end{table}

\begin{table}[htb!]
\footnotesize
\centering
\setlength{\tabcolsep}{0.58em}
\begin{tabular}{lllllllllll}
\hline
Feat.  &  \multicolumn{3}{l}{\textit{Weighted-F$_1$} (\%)}   &   &  & \\
\cline{2-11}
Size &  TF& TF-&  DTF- &  TF- & TF-&  TF-& STF-& TF-&  STF- & TF-\\
 &  & IDF&  IDF &  IDF & RF&  IGM& IGM& IGMi& IGMi & IDFC-RF\\
\hline
500 & 72.21 & 66.16 & 60.90 & 68.55 & 70.70 & 71.42 & 74.21 & 74.12 & 77.63 & \textbf{78.10} \\
1000 & 73.56 & 68.95 & 61.06 & 69.03 & 72.37 & 73.65 & 75.72 & 75.40 & 78.18 & \textbf{78.74} \\
2000 & 73.64 & 73.25 & 61.38 & 67.35 & 72.85 & 74.28 & 77.55 & 75.24 & 78.42 & \textbf{78.74} \\
4000 & 73.72 & 75.63 & 61.94 & 66.72 & 74.04 & 74.84 & 78.10 & 75.40 & 78.10 & \textbf{79.14} \\
6000 & 73.64 & 76.51 & 61.54 & 67.91 & 74.36 & 75.40 & 77.94 & 75.64 & 78.26 & \textbf{78.98} \\
8000 & 73.80 & 76.35 & 61.22 & 69.35 & 74.44 & 75.48 & 77.78 & 75.56 & 78.74 & \textbf{78.98} \\
10000 & 74.04 & 76.11 & 61.38 & 71.10 & 74.20 & 75.48 & 78.02 & 75.56 & 79.46 & \textbf{79.85} \\
12000 & 73.88 & 76.51 & 61.46 & 71.81 & 74.12 & 74.84 & 78.18 & 75.32 & 79.46 & \textbf{79.77} \\
14000 & 74.60 & 76.27 & 61.38 & 71.57 & 75.08 & 75.72 & 78.74 & 75.87 & 79.30 & \textbf{79.93} \\

\hline
\end{tabular}
\caption{Performances of TF, TF-IDF, STF-IDF, TF-IDF-ICF, TF-RF, TF-IGM, STF-IGM, TF-IGM$_{imp}$, STF-IGM$_{imp}$ and TF-IDFC-RF  using SVM classifier on \textit{Sarcasm} dataset.}
\label{tab:sarcSMO}
\end{table}

\subsection{Performance comparisons on the Subjectivity dataset}
\begin{figure}[htb!]
\centering
    \centering
    \subfloat[NB]{\resizebox{0.58\textwidth}{!}{\input{fig4a.tex} }}%
    \subfloat[SVM]{\resizebox{0.42\textwidth}{!}{\input{fig4b.tex} }}%
\caption{Weighted-F$_1$ scores obtained using NB and SVM classifiers with $10$ term weighting schemes on \textit{Subjectivity} dataset.}
\label{fig:sub}
\end{figure}
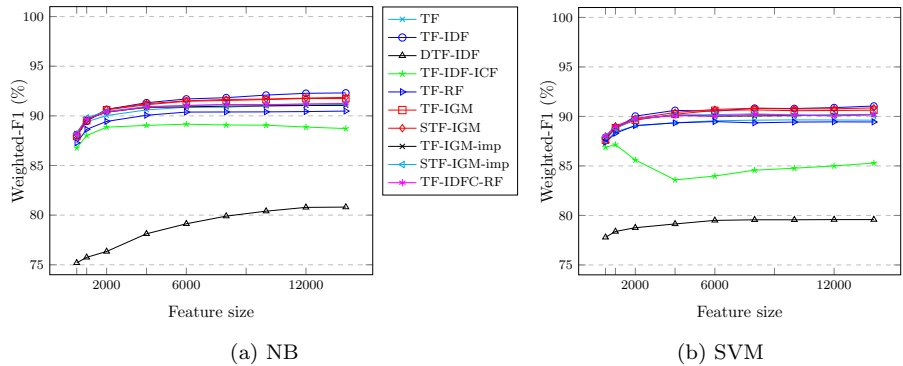

Subjectivity dataset is also a balanced dataset, however, it consists of $10,000$ sentences, as pointed out in Section \ref{subsec:sub}. Figures \ref{fig:sub}(a) and \ref{fig:sub}(b) present Weighted F$_1$ scores achieved on Subjectivity dataset. In most cases, F$_1$ scores obtained with TF-IDF surpassed the other term weighting schemes. 
Concerning SVM classifier, TF-IDFC-RF presents the best F$_1$ results when feature size is $500$ and $1000$. In most cases, F$_1$ scores achieved by the remaining weighting schemes with SVM increase as the feature sizes increase. DTF-IDF and TF-IDF-ICF show the most unsatisfactory results considering all feature sizes. TF-RF scheme also presents a poor result; however, its results can be much better than DTF-IDF and TF-IDF-ICF schemes.

Tables \ref{tab:subnb} and \ref{tab:subsvm} report the detailed F$_1$ results obtained with Subjectivity dataset. Except DTF-IDF scheme, in most cases, the weighting schemes presented better F$_1$ results considering the NB classifier. TF-IDF generally shows the best performance with both classifiers. Table \ref{tab:subnb} shows that TF-IGM$_{imp}$ and  STF-IGM$_{imp}$ achieved the best result when feature sizes are $500$ and $1,000$ respectively. 

As indicated in Table \ref{tab:subsvm}, TF-IDFC-RF performs better than TF-IDF when feature sizes are $500$ and $1,000$. This table also present that STF-IGM achieved the best result when the feature size is $6,000$.
The F$_1$ score difference between TF-IDF and TF-IDFC-RF is lower than $0.2$ percentage points in all feature sizes with SVM and NB. 



\begin{table}[htb!]
\footnotesize
\centering
\setlength{\tabcolsep}{0.58em}
\begin{tabular}{lllllllllll}
\hline
Feat.  &  \multicolumn{3}{l}{\textit{Weighted-F$_1$} (\%)}   &   &  & \\
\cline{2-11}
Size &  TF& TF-&  DTF- &  TF- & TF-&  TF-& STF-& TF-&  STF- & TF-\\
 &  & IDF&  IDF &  ICF & RF&  IGM& IGM& IGMi& IGMi & IDFC-RF \\
\hline
500 & 88.02 & 88.02 & 75.19 & 86.78 & 87.29 & 87.93 & 88.10 & \textbf{88.25} & 88.22 & 88.20 \\
1000 & 89.40 & 89.45 & 75.75 & 88.02 & 88.64 & 89.53 & 89.54 & 89.66 & \textbf{89.82} & 89.65 \\
2000 & 90.03 & \textbf{90.65} & 76.34 & 88.86 & 89.44 & 90.61 & \textbf{90.65} & 90.37 & 90.46 & 90.38 \\
4000 & 90.60 & \textbf{91.32} & 78.12 & 89.04 & 90.06 & 91.12 & 91.22 & 90.85 & 90.93 & 90.88 \\
6000 & 90.87 & \textbf{91.70} & 79.13 & 89.16 & 90.38 & 91.47 & 91.51 & 90.90 & 91.05 & 91.03 \\
8000 & 90.87 & \textbf{91.83} & 79.90 & 89.08 & 90.39 & 91.55 & 91.65 & 90.98 & 91.14 & 91.14 \\
10000 & 91.03 & \textbf{92.09} & 80.40 & 89.06 & 90.40 & 91.64 & 91.69 & 91.00 & 91.14 & 91.12 \\
12000 & 91.08 & \textbf{92.27} & 80.78 & 88.87 & 90.43 & 91.75 & 91.79 & 91.04 & 91.18 & 91.19 \\
14000 & 91.16 & \textbf{92.31} & 80.81 & 88.72 & 90.48 & 91.74 & 91.86 & 91.05 & 91.22 & 91.25 \\

\hline
\end{tabular}
\caption{Performances of TF, TF-IDF, DTF-IDF, TF-IDF-ICF, TF-RF, TF-IGM, STF-IGM, TF-IGM$_{imp}$, STF-IGM$_{imp}$ and TF-IDFC-RF using NB classifier on \textit{Subjectivity} dataset.}
\label{tab:subnb}
\end{table}

\begin{table}[htb!]
\footnotesize
\centering
\setlength{\tabcolsep}{0.58em}
\begin{tabular}{lllllllllll}
\hline
Feat.  &  \multicolumn{3}{l}{\textit{Weighted-F$_1$} (\%)}   &   &  & \\
\cline{2-11}
Size &  TF& TF-&  DTF- &  TF- & TF-&  TF-& STF-& TF-&  STF- & TF-\\
 &  & IDF&  IDF &  ICF & RF&  IGM& IGM& IGMi& IGMi & IDFC-RF\\
 \hline
500 & 87.49 & 87.84 & 77.79 & 86.88 & 72.23 & 87.61 & 87.99 & 87.88 & 88.00 & \textbf{88.03} \\
1000 & 88.45 & 88.73 & 78.38 & 87.15 & 74.20 & 88.85 & 89.04 & 88.86 & 88.77 & \textbf{88.92} \\
2000 & 89.04 & \textbf{90.03} & 78.76 & 85.60 & 75.31 & 89.62 & 89.79 & 89.72 & 89.73 & 89.76 \\
4000 & 89.36 & \textbf{90.61} & 79.14 & 83.59 & 75.89 & 90.22 & 90.40 & 90.14 & 90.15 & 90.07 \\
6000 & 89.56 & 90.53 & 79.50 & 83.98 & 75.69 & 90.61 & \textbf{90.71} & 90.03 & 90.18 & 90.04 \\
8000 & 89.62 & \textbf{90.83} & 79.56 & 84.57 & 75.99 & 90.66 & 90.82 & 90.05 & 90.26 & 90.20 \\
10000 & 89.65 & \textbf{90.80} & 79.56 & 84.77 & 75.89 & 90.58 & 90.76 & 90.08 & 90.16 & 90.13 \\
12000 & 89.64 & \textbf{90.89 }& 79.58 & 85.00 & 75.82 & 90.60 & 90.77 & 90.13 & 90.14 & 90.05 \\
14000 & 89.62 & \textbf{91.05} & 79.58 & 85.29 & 75.88 & 90.63 & 90.86 & 90.18 & 90.19 & 90.18 \\

\hline
\end{tabular}
\caption{Performances of TF,  TF-IDF,  DTF-IDF, TF-IDF-ICF,  TF-RF,  TF-IGM,  STF-IGM,  TF-IGM$_{imp}$,  STF-IGM$_{imp}$ and  TF-IDFC-RF using SVM classifier on \textit{Subjectivity} dataset.}
\label{tab:subsvm}
\end{table}
\subsection{Performance comparisons on the Movie Review dataset}
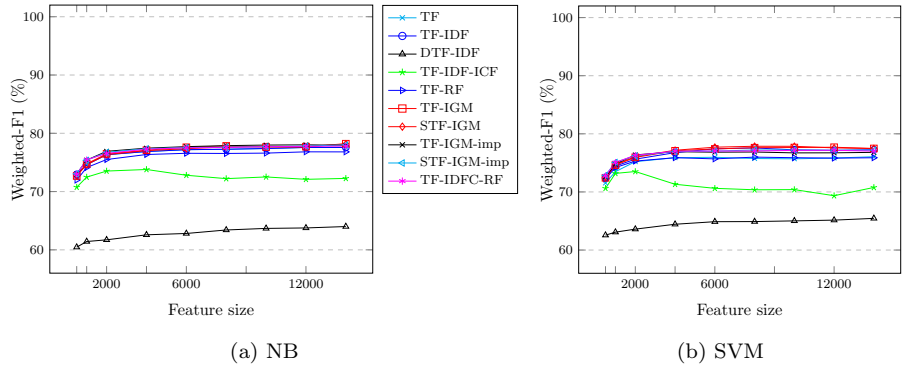
\begin{figure}[htb!]
\centering
    \centering
    \subfloat[NB]{\resizebox{0.58\textwidth}{!}{\input{fig5a.tex} }}%
    \subfloat[SVM]{\resizebox{0.42\textwidth}{!}{\input{fig5b.tex} }}%
\caption{Weighted-F$_1$ scores obtained using NB and SVM classifiers with $10$ term weighting schemes on \textit{Movie Review} dataset.}
\label{fig:mr}
\end{figure}

Figures~\ref{fig:mr}(a) and \ref{fig:mr}(b) present Weighted-F$_1$ scores achieved on Movie Review dataset.
This is a balanced dataset containing 10,662 movie-reviews ``snippets'', as indicated in Section \ref{subsec:mr}. One can note that IGM variations achieve the best performance, presenting values greater than 77\% for NB and SVM. The second-best scheme provided for SVM is TF-IDFC-RF, presenting F$_1$ values of more than 77\% when features size is greater than $2,000$. TF-IDF-ICF and DTF-IDF present the worst results with both classifiers. TF-RF achieve slightly better results than TF-IDF-ICF; however, it showed lower values than all the other schemes.

Tables \ref{tab:mrnb} and \ref{tab:mrsvm} show the F$_1$ values obtained for each term weighting scheme. Although generally, IGM schemes show the best results, it is possible to note in Table \ref{tab:mrnb} that TF-IDF-RF is the best weighting scheme when feature size is $1,000$, presenting an F$_1$ of $75.45$. In the remaining cases, IGM schemes performed better. However, it is important to emphasize that there is not an IGM scheme that is the best one for all feature sizes (they vary, sometimes is one, another time is another).

When considering the SVM classifier, one can note that IGM schemes outperformed all other weighting schemes. There is a small difference between TF-IDF and IGM schemes. The same small difference occurs between TF-IDFC-RF and IGM schemes. DTF-IDF presents the best F$_1$ of $64.00$, which is far from the $78.21$ F$_1$ achieved by TF-IGM.


\begin{table}[htb!]
\footnotesize
\centering
\setlength{\tabcolsep}{0.58em}
\begin{tabular}{lllllllllll}
\hline
Feat.  &  \multicolumn{3}{l}{\textit{Weighted-F$_1$} (\%)}   &   &  & \\
\cline{2-11}
Size &  TF& TF-&  DTF- &  TF- & TF-&  TF-& STF-& TF-&  STF- & TF-\\
 &  & IDF&  IDF &  ICF & RF&  IGM& IGM& IGMi& IGMi & IDFC-RF  \\
\hline
500 & 72.96 & 72.78 & 60.46 & 70.75 & 71.97 & 72.68 & 72.73 & 73.00 & \textbf{73.05} & 72.97 \\
1000 & 74.63 & 75.18 & 61.43 & 72.48 & 74.12 & 74.76 & 74.62 & 75.40 & 75.38 & \textbf{75.45} \\
2000 & 76.33 & 76.49 & 61.72 & 73.52 & 75.50 & 76.47 & 76.34 & \textbf{76.91} & 76.74 & 76.55 \\
4000 & 76.87 & 77.23 & 62.58 & 73.79 & 76.36 & 77.10 & 77.03 & \textbf{77.45} & 77.36 & 77.27 \\
6000 & 77.19 & 77.62 & 62.81 & 72.80 & 76.57 & 77.60 & 77.35 & \textbf{77.71} & 77.54 & 77.51 \\
8000 & 77.26 & 77.72 & 63.41 & 72.22 & 76.54 & 77.83 & 77.67 & \textbf{77.88} & 77.63 & 77.56 \\
10000 & 77.39 & 77.79 & 63.68 & 72.51 & 76.60 & 77.65 & 77.70 & \textbf{77.98} & 77.70 & 77.71 \\
12000 & 77.57 & 77.80 & 63.76 & 72.10 & 76.82 & 77.70 & 77.65 & \textbf{78.02} & 77.81 & 77.76 \\
14000 & 77.57 & 77.86 & 64.00 & 72.26 & 76.81 & \textbf{78.21} & 78.17 & 77.94 & 78.02 & 77.80 \\

\hline
\end{tabular}
\caption{Performances of TF,  TF-IDF,  DTF-IDF, TF-IDF-ICF,  TF-RF,  TF-IGM,  STF-IGM,  TF-IGM$_{imp}$,  STF-IGM$_{imp}$ and  TF-IDFC-RF using NB classifier on \textit{MR} dataset.}
\label{tab:mrnb}
\end{table}

\begin{table}[htb!]
\footnotesize
\centering
\setlength{\tabcolsep}{0.58em}
\begin{tabular}{lllllllllll}
\hline
Feat.  &  \multicolumn{3}{l}{\textit{Weighted-F$_1$} (\%)}   &   &  & \\
\cline{2-11}
Size &  TF& TF-&  DTF- &  TF- & TF-&  TF-& STF-& TF-&  STF- & TF-\\
 &  & IDF&  IDF &  ICF & RF&  IGM& IGM& IGMi& IGMi & IDFC-RF \\
\hline
500 & 71.54 & 72.43 & 62.56 & 70.62 & 72.23 & 72.43 & 72.55 & 72.59 & \textbf{72.87} & 72.77 \\
1000 & 73.49 & 74.45 & 63.08 & 73.19 & 74.20 & 74.63 & 74.70 & 74.81 & \textbf{75.10} & 75.08 \\
2000 & 75.21 & 75.64 & 63.60 & 73.51 & 75.31 & 75.95 & 76.07 & \textbf{76.37} & 76.22 & 76.30 \\
4000 & 75.89 & 76.76 & 64.44 & 71.32 & 75.89 & 77.08 & \textbf{77.17} & 76.92 & 76.97 & 77.01 \\
6000 & 75.96 & 77.42 & 64.88 & 70.62 & 75.69 & 77.40 & \textbf{77.70} & 76.84 & 77.02 & 77.04 \\
8000 & 75.75 & 77.57 & 64.90 & 70.37 & 75.99 & 77.61 & \textbf{77.86} & 76.87 & 77.22 & 77.05 \\
10000 & 75.70 & 77.24 & 65.02 & 70.40 & 75.89 & 77.64 & \textbf{77.85} & 76.72 & 77.21 & 77.15 \\
12000 & 75.83 & 77.14 & 65.15 & 69.35 & 75.82 & \textbf{77.69} & 77.58 & 76.73 & 77.18 & 77.17 \\
14000 & 76.06 & 77.27 & 65.45 & 70.77 & 75.88 & \textbf{77.48} & 77.46 & 76.81 & 77.19 & 77.02 \\

\hline
\end{tabular}
\caption{Performances of TF,  TF-IDF,  DTF-IDF, TF-IDF-ICF,  TF-RF,  TF-IGM,  STF-IGM,  TF-IGM$_{imp}$,  STF-IGM$_{imp}$ and  TF-IDFC-RF using SVM classifier on \textit{MR} dataset.}
\label{tab:mrsvm}
\end{table}

\subsection{Discussion}

The performance assessment of different term weighting schemes in classification tasks was executed with four two-class datasets. Results are generally better with SVM classifier considering two datasets (Polarity and Sarcasm) and better with NB on the other two datasets (Subjectivity and Movie Review). 
Prior work reported TF-IGM$_{imp}$ and STF-IGM$_{imp}$ generally outperform TF-IGM and STF-IGM as well as STF-IGM$_{imp}$ generally outperforms TF-IGM$_{imp}$ \cite{Dogan201945}. However, on all datasets considered in this work, this pattern is not found. For example, on the Polarity dataset, not always STF-IGM$_{imp}$ is better than STF-IGM. The same pattern occurs on the Subjectivity dataset. This is an important finding, since, as the authors know, this is the first study to conduct experiments with IGM based schemes and polarity datasets.

The results obtained in previous studies indicated good results with TF-RF \cite{Lan2009721, Chen20161339}. As reported in \cite{Chen20161339}, sometimes TF-RF outperforms TF-IGM and STF-IGM. However, our results revealed that TF-RF is almost always worse than all IGM based schemes with NB and SVM on all four datasets. On the other hand, TF-IDFC-RF presented results equal to or better than TF-RF in all experiments.

TF and TF-ICF did not achieve the best results considering all four datasets. TF-IDF presented the best results on the Subjectivity dataset, and DTF-IDF showed the best results with the Sarcasm dataset and NB classifier.

Our results provide compelling evidence that TF-IDFC-RF achieves better results than the other nine weighting schemes on two datasets with NB and SVM. It is important to note that when considering IGM schemes, no one scheme performs better on a specific dataset. When considering different feature sizes, one time, the best can be a squared IGM based scheme, another time a non-squared IGM scheme. Since there is only one version of TF-IDFC-RF, it presents more consistent results. 

\section{Conclusion and Future work}
\label{sec:conclusion}

In this work, we have proposed a novel supervised term weighting scheme named TF-IDFC-RF to be used in Sentiment Analysis tasks, more specifically, in the binary classification problem. The proposed scheme is based on two other schemes: TF-IDF and TF-RF.  TF-IDFC-RF is inspired by the fact that the IDF factor can be used for each class, referred to as the Inverse Document Frequency in Class (IDFC). On the other hand, since TF-RF has produced good results in the literature, we were also inspired TF-IDFC-RF on it.  The most important concept of TF-IDFC-RF is that it aims to consider intra-class and inter-class distribution to weight the terms. 

The performance of TF-IDFC-RF is compared with nine other term weighting schemes, including TF-IDF and TF-RF. These schemes also encompass the IGM based schemes, since they outperformed several other schemes in recent work \cite{Chen20161339, Dogan201945}.  It is important to stress that, as stated in \cite{Chen20161339}, TF-IGM schemes are especially suitable for multi-class text classification applications, however, they can be used for binary classification. SVM and NB classifiers were utilized to perform the experiments with different feature sizes.

The experiments show that TF-IDFC-RF outperforms all schemes with NB and SVM on two datasets. IGM based schemes achieved the best results in one dataset. Finally, TF-IDF presented performed better than all other schemes in one dataset. In future work, we will conduct comparative studies with TF-IDFC-RF in multi-class datasets. We also plan to produce experiments with larger datasets.

\section*{Acknowledgements}
\label{sec:ack}
This research was financed by the Coordena\c{c}\~ao de Aperfeiçoamento de Pessoal de N\'ivel Superior (CAPES) – Brazil. Financing Code: 001.
This research was partly funded by CNPq and FAPERJ.

\section*{References}

\bibliography{references}

\end{document}

%% file: fig2a.tex
\pgfplotsset{every tick label/.append style={font=\footnotesize}}
\begin{tikzpicture}
\begin{axis}[
    xlabel={Feature size},
    ylabel={Weighted-F1 (\%)},
    y label style={at={(.085,0.5)}},
    ymin=49, ymax=101,
    xtick={500,1000,2000,4000,6000, 8000,10000,12000,14000},
    xticklabels={,,2000,4000,6000,8000,10000,12000,14000},
    ytick={50, 60, 70,80,90,100},
    legend pos = outer north east,
     legend style={font=\footnotesize,mark options={scale=1.1}},
     legend cell align={left},
    ymajorgrids=true,
    grid style=dashed,
    scaled ticks=false,
]
 \addplot
[color=cyan, mark=x, semithick]
    coordinates {
    
(500,79.25)
(1000,80.35)
(2000,79.95)
(4000,78.4)
(6000,77.3)
(8000,76.15)
(10000,75.75)
(12000,74.9)
(14000,74.35)
    };
 \addplot
[color=blue, mark=o, semithick]
    coordinates {
(500,79.7)
(1000,81.35)
(2000,82.0)
(4000,81.95)
(6000,82.0)
(8000,82.2)
(10000,82.45)
(12000,81.9)
(14000,82.15)
    };
 \addplot
[color=black, mark=triangle, semithick]
    coordinates {
(500,55.45)
(1000,53.55)
(2000,51.85)
(4000,51.15)
(6000,50.7)
(8000,50.6)
(10000,50.35)
(12000,50.35)
(14000,50.4)
    };
   \addplot
[color=green, mark=star, semithick]  
  coordinates {
(500,72.3)
(1000,72.05)
(2000,73.9)
(4000,74.0)
(6000,74.65)
(8000,74.05)
(10000,75.05)
(12000,75.4)
(14000,75.25)
    };

       \addplot
[color=blue, mark=triangle, every mark/.append style={rotate=270}, semithick]
  coordinates {
(500,72.9)
(1000,70.35)
(2000,68.05)
(4000,67.45)
(6000,67.85)
(8000,67.25)
(10000,67.6)
(12000,67.75)
(14000,68.35)
    };
    \addplot
[color=red, mark=square, semithick]
    coordinates {
(500,81.6)
(1000,81.0)
(2000,80.75)
(4000,79.85)
(6000,79.2)
(8000,78.55)
(10000,78.0)
(12000,78.15)
(14000,78.0)
    };
 \addplot [color=red, mark=diamond, semithick]
    coordinates {
(500,84.1)
(1000,84.35)
(2000,84.25)
(4000,84.8)
(6000,84.3)
(8000,83.5)
(10000,83.6)
(12000,83.65)
(14000,83.65)
    };
 \addplot [color=black, mark=x, semithick]
    coordinates {
(500,81.3)
(1000,81.9)
(2000,81.45)
(4000,80.15)
(6000,79.85)
(8000,79.1)
(10000,78.8)
(12000,78.8)
(14000,78.5)
    };
 \addplot [color=cyan, mark=triangle, every mark/.append style={rotate=90}, semithick]
    coordinates {
(500,84.05)
(1000,84.6)
(2000,84.75)
(4000,83.9)
(6000,83.75)
(8000,83.45)
(10000,83.45)
(12000,83.7)
(14000,83.65)
    };
   \addplot
   [color=pinkk, mark=asterisk, semithick]
  coordinates {
(500,83.85)
(1000,85.20)
(2000,85.40)
(4000,85.10)
(6000,84.25)
(8000,84.30)
(10000,84.30)
(12000,84.30)
(14000,84.10)
    };  
    
\legend{TF, TF-IDF, DTF-IDF, TF-IDF-ICF, TF-RF, TF-IGM, STF-IGM, TF-IGM-imp, STF-IGM-imp, TF-IDFC-RF}

\end{axis}
\end{tikzpicture}

%% file: fig2b.tex
\pgfplotsset{every tick label/.append style={font=\footnotesize}}
\begin{tikzpicture}
\begin{axis}[
    xlabel={Feature size},
    ylabel={Weighted-F1 (\%)},
    y label style={at={(.085,0.5)}},
    ymin=49, ymax=101,
    xtick={500,1000,2000,4000,6000, 8000,10000,12000,14000},
    xticklabels={,,2000,4000,6000,8000,10000,12000,14000},
    ytick={50, 60, 70,80,90,100},
    legend pos = outer north east,
     legend style={font=\footnotesize,mark options={scale=1.1}},
     legend cell align={left},
    ymajorgrids=true,
    grid style=dashed,
    scaled ticks=false,
]
 \addplot
[color=cyan, mark=x, semithick]
    coordinates {
(500,75.5)
(1000,75.65)
(2000,77.5)
(4000,77.7)
(6000,78.0)
(8000,78.25)
(10000,78.15)
(12000,78.3)
(14000,78.2)
    };
 \addplot
[color=blue, mark=o, semithick]
    coordinates {
(500,79.15)
(1000,81.45)
(2000,81.95)
(4000,82.95)
(6000,83.25)
(8000,83.75)
(10000,83.45)
(12000,83.85)
(14000,84.25)
    };
 \addplot
[color=black, mark=triangle, semithick]
    coordinates {
(500,66.95)
(1000,67.3)
(2000,67.2)
(4000,67.35)
(6000,67.6)
(8000,67.4)
(10000,67.6)
(12000,67.85)
(14000,67.85)
    };
   \addplot
[color=green, mark=star, semithick]   
  coordinates {
(500,74.9)
(1000,72.15)
(2000,78.0)
(4000,80.5)
(6000,81.8)
(8000,82.1)
(10000,82.15)
(12000,83.15)
(14000,82.75)
    };

       \addplot
[color=blue, mark=triangle, every mark/.append style={rotate=270}, semithick]
  coordinates {
(500,75.25)
(1000,76.45)
(2000,77.3)
(4000,78.85)
(6000,78.95)
(8000,78.8)
(10000,79.15)
(12000,79.05)
(14000,79.4)
    };   
    \addplot
[color=red, mark=square, semithick]
    coordinates {
(500,77.35)
(1000,78.7)
(2000,79.9)
(4000,81.25)
(6000,81.65)
(8000,81.45)
(10000,81.6)
(12000,81.35)
(14000,81.75)
    };
 \addplot [color=red, mark=diamond, semithick]
    coordinates {
 (500,83.45)
(1000,84.7)
(2000,85.4)
(4000,86.6)
(6000,86.85)
(8000,87.35)
(10000,87.65)
(12000,87.85)
(14000,87.85)

    };   
 \addplot [color=black, mark=x, semithick]
    coordinates {
(500,77.35)
(1000,79.0)
(2000,80.35)
(4000,81.25)
(6000,81.4)
(8000,81.95)
(10000,81.85)
(12000,81.75)
(14000,81.65)
    };   
 \addplot [color=cyan, mark=triangle, every mark/.append style={rotate=90}, semithick]
    coordinates {
(500,83.8)
(1000,84.75)
(2000,85.9)
(4000,87.1)
(6000,87.65)
(8000,88.1)
(10000,87.7)
(12000,87.65)
(14000,87.85)
    };   
        
   \addplot
   [color=pinkk, mark=asterisk, semithick]
  coordinates {
(500,84.65)
(1000,86.05)
(2000,87.30)
(4000,87.70)
(6000,87.45)
(8000,87.70)
(10000,87.75)
(12000,88.05)
(14000,87.90)
    };     

\end{axis}
\end{tikzpicture}

%% file: fig3a.tex
\pgfplotsset{every tick label/.append style={font=\footnotesize}}
\begin{tikzpicture}
\begin{axis}[
    xlabel={Feature size},
    ylabel={Weighted-F1 (\%)},
    y label style={at={(.085,0.5)}},
    ymin=49, ymax=101,
    xtick={500,1000,2000,4000,6000, 9000,12000,16000,20000,25000},
    xticklabels={,,2000,,6000,,12000,16000,20000,25000},
    ytick={50, 60, 70, 80,90,100},
    legend pos = outer north east,
     legend style={font=\footnotesize,mark options={scale=1.1}},
     legend cell align={left},
    ymajorgrids=true,
    grid style=dashed,
    scaled ticks=false,
]
 \addplot
[color=cyan, mark=x, semithick]
    coordinates {
(500,65.21)
(1000,65.13)
(2000,65.13)
(4000,65.13)
(6000,65.13)
(8000,65.13)
(10000,65.13)
(12000,65.13)
(14000,65.13)
    };   
 \addplot
[color=blue, mark=o, semithick]
    coordinates {
(500,65.29)
(1000,65.37)
(2000,65.37)
(4000,65.29)
(6000,65.29)
(8000,65.21)
(10000,65.13)
(12000,65.13)
(14000,65.13)
    };   
 \addplot
[color=black, mark=triangle, semithick]
    coordinates {
(500,65.76)
(1000,67.36)
(2000,66.72)
(4000,66.32)
(6000,66.88)
(8000,66.64)
(10000,66.56)
(12000,66.48)
(14000,66.8)
    };   
   \addplot
[color=green, mark=star, semithick]   
  coordinates {
(500,64.73)
(1000,64.41)
(2000,62.9)
(4000,61.14)
(6000,62.58)
(8000,62.02)
(10000,62.26)
(12000,61.62)
(14000,60.91)
    };   

       \addplot
[color=blue, mark=triangle, every mark/.append style={rotate=270}, semithick]   
  coordinates {
(500,65.05)
(1000,65.29)
(2000,65.21)
(4000,65.21)
(6000,65.13)
(8000,65.13)
(10000,65.13)
(12000,65.13)
(14000,65.13)
    };   
    \addplot
[color=red, mark=square, semithick]
    coordinates {
(500,65.29)
(1000,65.21)
(2000,65.13)
(4000,65.13)
(6000,65.13)
(8000,65.13)
(10000,65.13)
(12000,65.13)
(14000,65.13)
    };   
 \addplot [color=red, mark=diamond, semithick]
    coordinates {
(500,65.21)
(1000,65.37)
(2000,65.13)
(4000,65.13)
(6000,65.13)
(8000,65.13)
(10000,65.13)
(12000,65.13)
(14000,65.13)
    };   
 \addplot [color=black, mark=x, semithick]
    coordinates {
(500,65.21)
(1000,65.21)
(2000,65.21)
(4000,65.13)
(6000,65.13)
(8000,65.13)
(10000,65.13)
(12000,65.13)
(14000,65.13)
    };     
 \addplot [color=cyan, mark=triangle, every mark/.append style={rotate=90}, semithick]
    coordinates {
(500,65.21)
(1000,65.21)
(2000,65.37)
(4000,65.13)
(6000,65.13)
(8000,65.13)
(10000,65.13)
(12000,65.13)
(14000,65.13)
    };   
        
   \addplot
   [color=pinkk, mark=asterisk, semithick]
  coordinates {
(500,66.96)
(1000,65.76)
(2000,65.37)
(4000,65.29)
(6000,65.13)
(8000,65.13)
(10000,65.13)
(12000,65.13)
(14000,65.13)
    };       
    
\legend{TF, TF-IDF, DTF-IDF, TF-IDF-ICF, TF-RF, TF-IGM, STF-IGM, TF-IGM-imp, STF-IGM-imp, TF-IDFC-RF}

\end{axis}
\end{tikzpicture}

%% file: fig3b.tex
`\pgfplotsset{every tick label/.append style={font=\footnotesize}}
\begin{tikzpicture}
\begin{axis}[
    xlabel={Feature size},
    ylabel={Weighted-F1 (\%)},
    y label style={at={(.085,0.5)}},
    ymin=49, ymax=101,
    xtick={500,1000,2000,4000,6000, 9000,12000,16000,20000,25000},
    xticklabels={,,2000,,6000,,12000,16000,20000,25000},
    ytick={50, 60, 70, 80,90,100},
    legend pos = outer north east,
     legend style={font=\footnotesize,mark options={scale=1.1}},
     legend cell align={left},
    ymajorgrids=true,
    grid style=dashed,
    scaled ticks=false,
]

 \addplot
[color=cyan, mark=x, semithick]
    coordinates {
(500,72.21)
(1000,73.56)
(2000,73.64)
(4000,73.72)
(6000,73.64)
(8000,73.8)
(10000,74.04)
(12000,73.88)
(14000,74.6)
    };
 \addplot
[color=blue, mark=o, semithick]
    coordinates {
(500,66.16)
(1000,68.95)
(2000,73.25)
(4000,75.63)
(6000,76.51)
(8000,76.35)
(10000,76.11)
(12000,76.51)
(14000,76.27)
    };
 \addplot
[color=black, mark=triangle, semithick]
    coordinates {
(500,60.9)
(1000,61.06)
(2000,61.38)
(4000,61.94)
(6000,61.54)
(8000,61.22)
(10000,61.38)
(12000,61.46)
(14000,61.38)
    };
   \addplot
[color=green, mark=star, semithick]  
    coordinates {
(500,68.55)
(1000,69.03)
(2000,67.35)
(4000,66.72)
(6000,67.91)
(8000,69.35)
(10000,71.1)
(12000,71.81)
(14000,71.57)
    };

       \addplot
[color=blue, mark=triangle, every mark/.append style={rotate=270}, semithick]
    coordinates {
(500,70.7)
(1000,72.37)
(2000,72.85)
(4000,74.04)
(6000,74.36)
(8000,74.44)
(10000,74.2)
(12000,74.12)
(14000,75.08)
    };
    \addplot
[color=red, mark=square, semithick]
    coordinates {
(500,71.42)
(1000,73.65)
(2000,74.28)
(4000,74.84)
(6000,75.4)
(8000,75.48)
(10000,75.48)
(12000,74.84)
(14000,75.72)
    };
 \addplot [color=red, mark=diamond, semithick]
    coordinates {
(500,74.21)
(1000,75.72)
(2000,77.55)
(4000,78.1)
(6000,77.94)
(8000,77.78)
(10000,78.02)
(12000,78.18)
(14000,78.74)
    };
 \addplot [color=black, mark=x, semithick]
    coordinates {
(500,74.12)
(1000,75.4)
(2000,75.24)
(4000,75.4)
(6000,75.64)
(8000,75.56)
(10000,75.56)
(12000,75.32)
(14000,75.87)
    };
 \addplot [color=cyan, mark=triangle, every mark/.append style={rotate=90}, semithick]
    coordinates {
(500,77.63)
(1000,78.18)
(2000,78.42)
(4000,78.1)
(6000,78.26)
(8000,78.74)
(10000,79.46)
(12000,79.46)
(14000,79.3)
    }; 
        
   \addplot
   [color=pinkk, mark=asterisk, semithick]
    coordinates {
(500,78.1)
(1000,78.74)
(2000,78.74)
(4000,79.14)
(6000,78.98)
(8000,78.98)
(10000,79.85)
(12000,79.77)
(14000,79.93)
    };   
    
\end{axis}
\end{tikzpicture}

%% file: fig4a.tex
\pgfplotsset{every tick label/.append style={font=\footnotesize}}
\begin{tikzpicture}
\begin{axis}[
    xlabel={Feature size},
    ylabel={Weighted-F1 (\%)},
    y label style={at={(.085,0.5)}},
    ymin=74, ymax=101,
    xtick={500,1000,2000,4000,6000, 9000,12000,16000,20000,25000},
    xticklabels={,,2000,,6000,,12000,16000,20000,25000},
    ytick={75,80, 85, 90, 95, 100},
    legend pos = outer north east,
     legend style={font=\footnotesize,mark options={scale=1.1}},
     legend cell align={left},
    ymajorgrids=true,
    grid style=dashed,
    scaled ticks=false,
]
 \addplot
[color=cyan, mark=x, semithick]
    coordinates {
(500,88.02)
(1000,89.4)
(2000,90.03)
(4000,90.6)
(6000,90.87)
(8000,90.87)
(10000,91.03)
(12000,91.08)
(14000,91.16)
    };   
 \addplot
[color=blue, mark=o, semithick]
    coordinates {
(500,88.02)
(1000,89.45)
(2000,90.65)
(4000,91.32)
(6000,91.7)
(8000,91.83)
(10000,92.09)
(12000,92.27)
(14000,92.31)
    };    
 \addplot
[color=black, mark=triangle, semithick]
    coordinates {
(500,75.19)
(1000,75.75)
(2000,76.34)
(4000,78.12)
(6000,79.13)
(8000,79.9)
(10000,80.4)
(12000,80.78)
(14000,80.81)
    };     
   \addplot
[color=green, mark=star, semithick]   
  coordinates {
(500,86.78)
(1000,88.02)
(2000,88.86)
(4000,89.04)
(6000,89.16)
(8000,89.08)
(10000,89.06)
(12000,88.87)
(14000,88.72)
    };    

       \addplot
[color=blue, mark=triangle, every mark/.append style={rotate=270}, semithick]
  coordinates {
(500,87.29)
(1000,88.64)
(2000,89.44)
(4000,90.06)
(6000,90.38)
(8000,90.39)
(10000,90.4)
(12000,90.43)
(14000,90.48)
    };     
    \addplot
[color=red, mark=square, semithick]
    coordinates {
(500,87.93)
(1000,89.53)
(2000,90.61)
(4000,91.12)
(6000,91.47)
(8000,91.55)
(10000,91.64)
(12000,91.75)
(14000,91.74)
    };   
 \addplot [color=red, mark=diamond, semithick]
    coordinates {
(500,88.1)
(1000,89.54)
(2000,90.65)
(4000,91.22)
(6000,91.51)
(8000,91.65)
(10000,91.69)
(12000,91.79)
(14000,91.86)
    };    
 \addplot [color=black, mark=x, semithick]
    coordinates {
(500,88.25)
(1000,89.66)
(2000,90.37)
(4000,90.85)
(6000,90.9)
(8000,90.98)
(10000,91.0)
(12000,91.04)
(14000,91.05)
    };     
 \addplot [color=cyan, mark=triangle, every mark/.append style={rotate=90}, semithick]
    coordinates {
(500,88.22)
(1000,89.82)
(2000,90.46)
(4000,90.93)
(6000,91.05)
(8000,91.14)
(10000,91.14)
(12000,91.18)
(14000,91.22)
    };   
        
   \addplot
   [color=pinkk, mark=asterisk, semithick]
  coordinates {
(500,88.2)
(1000,89.65)
(2000,90.38)
(4000,90.88)
(6000,91.03)
(8000,91.14)
(10000,91.12)
(12000,91.19)
(14000,91.25)
    };       
    
\legend{TF, TF-IDF, DTF-IDF, TF-IDF-ICF, TF-RF, TF-IGM, STF-IGM, TF-IGM-imp, STF-IGM-imp, TF-IDFC-RF}

\end{axis}
\end{tikzpicture}

%% file: fig4b.tex
\pgfplotsset{every tick label/.append style={font=\footnotesize}}
\begin{tikzpicture}
\begin{axis}[
    xlabel={Feature size},
    ylabel={Weighted-F1 (\%)},
    y label style={at={(.085,0.5)}},
    ymin=74, ymax=101,
    xtick={500,1000,2000,4000,6000, 9000,12000,16000,20000,25000},
    xticklabels={,,2000,,6000,,12000,16000,20000,25000},
    ytick={75, 80, 85, 90, 95, 100},
    legend pos = outer north east,
     legend style={font=\footnotesize,mark options={scale=1.1}},
     legend cell align={left},
    ymajorgrids=true,
    grid style=dashed,
    scaled ticks=false,
]
 \addplot
[color=cyan, mark=x, semithick]
    coordinates {
(500,87.49)
(1000,88.45)
(2000,89.04)
(4000,89.36)
(6000,89.56)
(8000,89.62)
(10000,89.65)
(12000,89.64)
(14000,89.62)
    };    
 \addplot
[color=blue, mark=o, semithick]
    coordinates {
(500,87.84)
(1000,88.73)
(2000,90.03)
(4000,90.61)
(6000,90.53)
(8000,90.83)
(10000,90.8)
(12000,90.89)
(14000,91.05)
    };    
 \addplot
[color=black, mark=triangle, semithick]
    coordinates {
(500,77.79)
(1000,78.38)
(2000,78.76)
(4000,79.14)
(6000,79.5)
(8000,79.56)
(10000,79.56)
(12000,79.58)
(14000,79.58)
    };     
   \addplot
[color=green, mark=star, semithick]   
  coordinates {
(500,86.88)
(1000,87.15)
(2000,85.6)
(4000,83.59)
(6000,83.98)
(8000,84.57)
(10000,84.77)
(12000,85.0)
(14000,85.29)
    };    

       \addplot
[color=blue, mark=triangle, every mark/.append style={rotate=270}, semithick]
  coordinates {
(500,87.38)
(1000,88.3)
(2000,89.08)
(4000,89.35)
(6000,89.48)
(8000,89.35)
(10000,89.43)
(12000,89.45)
(14000,89.44)
    };     
    \addplot
[color=red, mark=square, semithick]
    coordinates {
(500,87.61)
(1000,88.85)
(2000,89.62)
(4000,90.22)
(6000,90.61)
(8000,90.66)
(10000,90.58)
(12000,90.6)
(14000,90.63)
    };   
 \addplot [color=red, mark=diamond, semithick]
    coordinates {
(500,87.99)
(1000,89.04)
(2000,89.79)
(4000,90.4)
(6000,90.71)
(8000,90.82)
(10000,90.76)
(12000,90.77)
(14000,90.86)
    };    
 \addplot [color=black, mark=x, semithick]
    coordinates {
(500,87.88)
(1000,88.86)
(2000,89.72)
(4000,90.14)
(6000,90.03)
(8000,90.05)
(10000,90.08)
(12000,90.13)
(14000,90.18)
    };     
 \addplot [color=cyan, mark=triangle, every mark/.append style={rotate=90}, semithick]
    coordinates {
(500,88.0)
(1000,88.77)
(2000,89.73)
(4000,90.15)
(6000,90.18)
(8000,90.26)
(10000,90.16)
(12000,90.14)
(14000,90.19)
    };   
        
   \addplot
   [color=pinkk, mark=asterisk, semithick]
  coordinates {
(500,88.03)
(1000,88.92)
(2000,89.76)
(4000,90.07)
(6000,90.04)
(8000,90.2)
(10000,90.13)
(12000,90.05)
(14000,90.18)
    };       
    
\end{axis}
\end{tikzpicture}

%% file: fig5a.tex
\pgfplotsset{every tick label/.append style={font=\footnotesize}}
\begin{tikzpicture}
\begin{axis}[
    xlabel={Feature size},
    ylabel={Weighted-F1 (\%)},
    y label style={at={(.085,0.5)}},
    ymin=56, ymax=102,
    xtick={500,1000,2000,4000,6000, 9000,12000,16000,20000,25000},
    xticklabels={,,2000,,6000,,12000,16000,20000,25000},
    ytick={50, 60, 70, 80,90,100},
    legend pos = outer north east,
     legend style={font=\footnotesize,mark options={scale=1.1}},
     legend cell align={left},
    ymajorgrids=true,
    grid style=dashed,
    scaled ticks=false,
]
 \addplot
[color=cyan, mark=x, semithick]
    coordinates {
(500,72.96)
(1000,74.63)
(2000,76.33)
(4000,76.87)
(6000,77.19)
(8000,77.26)
(10000,77.39)
(12000,77.57)
(14000,77.57)
    };   
 \addplot
[color=blue, mark=o, semithick]
    coordinates {
(500,72.96)
(1000,74.63)
(2000,76.33)
(4000,76.87)
(6000,77.19)
(8000,77.26)
(10000,77.39)
(12000,77.57)
(14000,77.57)
    };   
 \addplot
[color=black, mark=triangle, semithick]
    coordinates {
(500,60.46)
(1000,61.43)
(2000,61.72)
(4000,62.58)
(6000,62.81)
(8000,63.41)
(10000,63.68)
(12000,63.76)
(14000,64.0)
    };   
   \addplot
[color=green, mark=star, semithick]   
  coordinates {
(500,70.75)
(1000,72.48)
(2000,73.52)
(4000,73.79)
(6000,72.8)
(8000,72.22)
(10000,72.51)
(12000,72.1)
(14000,72.26)
    };   

       \addplot
[color=blue, mark=triangle, every mark/.append style={rotate=270}, semithick]
  coordinates {
(500,71.97)
(1000,74.12)
(2000,75.5)
(4000,76.36)
(6000,76.57)
(8000,76.54)
(10000,76.6)
(12000,76.82)
(14000,76.81)
    };   
    \addplot
[color=red, mark=square, semithick]
    coordinates {
(500,72.68)
(1000,74.76)
(2000,76.47)
(4000,77.1)
(6000,77.6)
(8000,77.83)
(10000,77.65)
(12000,77.7)
(14000,78.21)
    };   
 \addplot [color=red, mark=diamond, semithick]
    coordinates {
(500,72.73)
(1000,74.62)
(2000,76.34)
(4000,77.03)
(6000,77.35)
(8000,77.67)
(10000,77.7)
(12000,77.65)
(14000,78.17)
    };   
 \addplot [color=black, mark=x, semithick]
    coordinates {
(500,73.0)
(1000,75.4)
(2000,76.91)
(4000,77.45)
(6000,77.71)
(8000,77.88)
(10000,77.98)
(12000,78.02)
(14000,77.94)
    };     
 \addplot [color=cyan, mark=triangle, every mark/.append style={rotate=90}, semithick]
    coordinates {
(500,73.05)
(1000,75.38)
(2000,76.74)
(4000,77.36)
(6000,77.54)
(8000,77.63)
(10000,77.7)
(12000,77.81)
(14000,78.02)
    };   
        
   \addplot
   [color=pinkk, mark=asterisk, semithick]
  coordinates {
(500,72.97)
(1000,75.45)
(2000,76.55)
(4000,77.27)
(6000,77.51)
(8000,77.56)
(10000,77.71)
(12000,77.76)
(14000,77.8)
    };       
    
\legend{TF, TF-IDF, DTF-IDF, TF-IDF-ICF, TF-RF, TF-IGM, STF-IGM, TF-IGM-imp, STF-IGM-imp, TF-IDFC-RF}

\end{axis}
\end{tikzpicture}

%% file: fig5b.tex
\pgfplotsset{every tick label/.append style={font=\footnotesize}}
\begin{tikzpicture}
\begin{axis}[
    xlabel={Feature size},
    ylabel={Weighted-F1 (\%)},
    y label style={at={(.085,0.5)}},
    ymin=56, ymax=102,
    xtick={500,1000,2000,4000,6000, 9000,12000,16000,20000,25000},
    xticklabels={,,2000,,6000,,12000,16000,20000,25000},
    ytick={60, 70, 80,90,100},
    legend pos = outer north east,
     legend style={font=\footnotesize,mark options={scale=1.1}},
     legend cell align={left},
    ymajorgrids=true,
    grid style=dashed,
    scaled ticks=false,
]

 \addplot
[color=cyan, mark=x, semithick]
    coordinates {
(500,71.54)
(1000,73.49)
(2000,75.21)
(4000,75.89)
(6000,75.96)
(8000,75.75)
(10000,75.7)
(12000,75.83)
(14000,76.06)
    };
 \addplot
[color=blue, mark=o, semithick]
    coordinates {
(500,72.43)
(1000,74.45)
(2000,75.64)
(4000,76.76)
(6000,77.42)
(8000,77.57)
(10000,77.24)
(12000,77.14)
(14000,77.27)
    };
 \addplot
[color=black, mark=triangle, semithick]
    coordinates {
(500,62.56)
(1000,63.08)
(2000,63.6)
(4000,64.44)
(6000,64.88)
(8000,64.9)
(10000,65.02)
(12000,65.15)
(14000,65.45)
    };
   \addplot
[color=green, mark=star, semithick]   
    coordinates {
(500,70.62)
(1000,73.19)
(2000,73.51)
(4000,71.32)
(6000,70.62)
(8000,70.37)
(10000,70.4)
(12000,69.35)
(14000,70.77)
    };

       \addplot
[color=blue, mark=triangle, every mark/.append style={rotate=270}, semithick]
    coordinates {
(500,72.23)
(1000,74.2)
(2000,75.31)
(4000,75.89)
(6000,75.69)
(8000,75.99)
(10000,75.89)
(12000,75.82)
(14000,75.88)
    };
    \addplot
[color=red, mark=square, semithick]
    coordinates {
(500,72.43)
(1000,74.63)
(2000,75.95)
(4000,77.08)
(6000,77.4)
(8000,77.61)
(10000,77.64)
(12000,77.69)
(14000,77.48)
    };
 \addplot [color=red, mark=diamond, semithick]
    coordinates {
(500,72.55)
(1000,74.7)
(2000,76.07)
(4000,77.17)
(6000,77.7)
(8000,77.86)
(10000,77.85)
(12000,77.58)
(14000,77.46)
    };
 \addplot [color=black, mark=x, semithick]
    coordinates {
(500,72.59)
(1000,74.81)
(2000,76.37)
(4000,76.92)
(6000,76.84)
(8000,76.87)
(10000,76.72)
(12000,76.73)
(14000,76.81)
    };
 \addplot [color=cyan, mark=triangle, every mark/.append style={rotate=90}, semithick]
    coordinates {
(500,72.87)
(1000,75.1)
(2000,76.22)
(4000,76.97)
(6000,77.02)
(8000,77.22)
(10000,77.21)
(12000,77.18)
(14000,77.19)
    }; 
        
   \addplot
   [color=pinkk, mark=asterisk, semithick]
    coordinates {
(500,72.77)
(1000,75.08)
(2000,76.3)
(4000,77.01)
(6000,77.04)
(8000,77.05)
(10000,77.15)
(12000,77.17)
(14000,77.02)
    };

\end{axis}
\end{tikzpicture}